# Deep Learning-Based Extraction of Promising Material Groups and Common Features from High-Dimensional Data: A Case of Optical Spectra of Inorganic Crystals

*Akira Takahashi*[1,2,*], *Yu Kumagai*[3], *Arata Takamatsu*[2], *and Fumiyasu Oba*[1,2,4]

[1]Materials and Structures Laboratory, Institute of Integrated Research, Institute of Science Tokyo
4259 Nagatsuta, Midori-ku, Yokohama 226-8501, Japan
[2]Laboratory for Materials and Structures, Institute of Innovative Research, Tokyo Institute of Technology
4259 Nagatsuta, Midori-ku, Yokohama 226-8501, Japan
[3]Institute for Materials Research, Tohoku University
2-2-1 Katahira, Aoba-ku, Sendai 980-8577, Japan
[4]Kanagawa Institute of Industrial Science and Technology (KISTEC)
705-1 Shimoimaizumi, Ebina 243-0435, Japan

[*]E-mail: takahashi.a.f9db@m.isct.ac.jp

**Abstract**

We report an interpretation method for deep learning models that allows us to handle high-dimensional spectral data in materials science. The proposed method uses feature extraction and clustering analysis to categorize materials into classes based on similarities in both spectral data and chemical characteristics such as elemental composition and atomic arrangement. As a demonstration, we apply this method to an atomistic line graph neural network (ALIGNN) model trained on first-principles calculation data of 2,681 metal oxides, chalcogenides, and related compounds for optical absorption spectrum prediction. Our analysis reveals key elemental species and their coordination environments that influence optical absorption onset characteristics. The method proposed herein is broadly applicable to the classification and interpretation of diverse spectral data, extending beyond the optical absorption spectra of inorganic crystals.

## 1. Introduction

The application of machine learning to materials science has gained significant attention in recent years, as exemplified by rapidly predicting various properties of solids such as band gap energies[1–3], melting points,[4] dielectric constants,[5–9] point-defect formation energies,[10–15] and electron affinities and ionization potentials.[16] Such machine learning prediction models have been demonstrated to accelerate and automate the identification and discovery of materials with desired properties.[15,17–34] Moreover, various machine learning interatomic potential methods have been proposed[35–45] and applied to predict stable structures, high-temperature and high-pressure phases, thermal and diffusion properties, and defect formation energies, as well as to perform molecular dynamics simulations and crystal structure generation.[46–68]

By applying explainable and interpretable artificial intelligence (AI) methods, such as feature importance analysis, Shapley additive explanations[69], local interpretable model-agnostic explanations[70], compressed sensing, symbolic regression, and graph visualization techniques, to prediction models based on hand-crafted descriptors and graph neural networks that directly predict target properties, it becomes possible to perform comprehensive analyses of databases.[6,7,71–84] Various interpretability methods have been proposed for machine learning potentials as well.[85–88]

Spectral data such as dielectric functions, optical spectra (absorption, reflection, and emission), and electronic and phononic densities of states are crucial for understanding and designing materials. Furthermore, there is a need to predict the dependence of a certain physical or chemical property on various parameters such as temperature, pressure, volume, and chemical composition. Although research on spectral prediction for solids has been relatively limited compared to scalar properties due to the high-dimensional output data, recent studies have demonstrated the construction of deep learning prediction models. Examples include the prediction model for electron energy-loss near edge structure (ELNES) and x-ray absorption near edge structure (XANES) spectra of specific metal oxides.[89] The prediction models of phonon spectra,[90,91] electronic densities of states,[92,93] and dielectric functions[94–96] have been constructed for general inorganic crystals, enabling predictions across different elemental compositions and crystal structures. Similar prediction models have also been developed for molecules, including prediction models of ELNES/XANES spectra[97] and overlap population diagrams.[98] Despite the importance of spectral data and the development of various prediction models for solids, the methodology for the interpretation of such prediction models of high-dimensional data is still limited.

This paper reports a methodology for extracting promising material groups and their common features through material classification using feature extraction and clustering of high-dimensional spectral data. We validate the proposed methodology using optical absorption spectra of metal oxides, chalcogenides, and related compounds as target property and systems. The optical absorption plays an important role in various applications. For example, the absorption characteristics of materials have a significant impact on their visual appearance, which is crucial for applications such as pigments, dyes, and functional optical coatings. Examples of typical inorganic pigments include metal oxides and chalcogenides such as zinc white (ZnO) and cadmium yellow (CdS).[99] Conversely, crystalline $SiO_2$ exhibit broad transparency across both visible and ultraviolet regions. This excellent transparency allows them to display vivid colors when doped with trace impurities, which are valuable as colorless hosts of gemstones. For example, amethyst is α-quartz containing $Fe^{3+}$ and other impurities that produce its characteristic purple coloration through charge-transfer transitions between $Fe^{3+}$ and $O^{2-}$, and when heated above 440°C, amethyst can transform into citrine, which displays yellow to orange hues due to changes in iron-related defect centers and the formation of interstitial $Fe^{3+}$ complexes.[100] Quartz is also used as an essential component in ultraviolet optical applications.[101,102] In addition, optical absorption properties of inorganic crystals crucially determine the performance of thin-film solar cells,[103,104] photocatalysts,[105] and photodetectors.[106]

This paper first addresses the construction of a first-principles calculation database for optical absorption spectra of the aforementioned inorganic crystals, and then confirms that atomistic line graph neural network (ALIGNN) prediction models[107] can be built with high accuracy. Using the constructed ALIGNN models, we perform material classification with the proposed methodology and discuss constituent elements and coordination environments commonly found in several material groups, exploring how these relate to existing materials science theories and practical applications.

## 2. Results and Discussion

### 2.1. Construction of the Dataset Using High-Throughput First-Principles Calculations

Theoretically, optical properties of solids including their absorption spectra can be obtained from first-principles calculations.[108–110] Recently, the development of high-throughput first-principles calculation methodologies and workflows enables us to evaluate absorption spectra of $10^3$–$10^4$ solids.[94,95,111] However, as Grunert et al. acknowledged,[95] previous studies about machine learning for the dielectric functions relied on the dataset by Kohn–Sham density functional theory calculations, which tend to severely underestimate band gaps[109,112,113] and thereby cause large errors in the onset of optical absorption spectra. To remedy this problem, it is necessary to use computationally demanding approaches such as hybrid functionals with the generalized Kohn–Sham scheme[114,115] and *GW* approximations based on many-body perturbation theory.[116] To make matters worse, optical absorption spectrum calculations for solids require very high-density *k*-point meshes when using band structure calculation methods, making it impractical to use such expensive approaches in high-throughput calculations.[117,118]

The modified Becke–Johnson (mBJ) potential[119–121] is known to offer band gaps better than standard local and semilocal density functionals with relatively low computational costs, and a dielectric functions dataset calculated using the mBJ potential is available in the JARVIS-DFT database.[122–124] Ibrahim and Ataca[96] developed a prediction model by leveraging transfer learning and multi-fidelity learning strategies[125–127] to integrate this data with extensive OptB88vdW functional calculations[121], which are known to systematically underestimate band gaps[122]. However, the mBJ potential tends to underestimate valence and conduction band widths[128], which would affect the shape of the dielectric functions.

Given this situation, we opted to construct an optical absorption spectrum database by combining spectrum calculations using the Perdew–Burke–Ernzerhof functional tuned for solids (PBEsol)[129] with Hubbard *U* corrections to localized states[130,131] and spectral onset corrections using non-self-consistent dielectric-dependent (nsc-dd) hybrid functional calculations[132] and the method suggested by Nishiwaki and Fujiwara,[118] as detailed in the Methods section. Here, a significant improvement in band gap prediction by the nsc-dd calculations over the standard PBEsol(+$U$) calculations has been demonstrated for a large number of semiconductors and insulators, including metal oxides and chalcogenides,[132–134] supporting the improved accuracy of the theoretical spectral onsets.

We retrieved 9,808 metal oxides, chalcogenides, and related substances such as carbonates, sulfates, phosphates, oxide- or chalcogenide-based mixed-anion compounds, and so forth from the Materials Project database[135] and constructed the computational database following the workflow shown in **Figure 1**. As can be seen, the computational workflow is highly complex, making it challenging to perform high-throughput calculations while maintaining consistency and efficiently utilizing computational resources. We enabled this using in-house programs, which rely on the PYMATGEN,[136] FIREWORKS,[137] CUSTODIAN,[136] ATOMATE[138], and VISE[12] codes. All the first-principles calculations were performed using the VASP code.[139,140]

This workflow yields optical absorption spectra and compound formation energies from PBEsol(+$U$) calculations, as well as band structures from both nsc-dd hybrid functional and PBEsol(+$U$) calculations. We adopted only materials that are stable with respect to competing phases within the in-house database in terms of the obtained formation energies with a plane-wave cutoff of 520 eV and have PBEsol(+$U$) band gaps wider than 0.3 eV for the machine learning dataset. The optical absorption coefficient was calculated from the dielectric function with $k$-mesh spacing of less than 0.133 Å$^{-1}$ and corrected using the Nishiwaki–Fujiwara method, employing the gap difference between nsc-dd hybrid and PBEsol(+$U$) calculations at the $k$-point that gives the minimum gap in the nsc-dd hybrid functional band structure calculations. Here, we define $E_{\mathrm{g}}^{\mathrm{PBEsol}(+U),\mathrm{band}}$ as the corresponding PBEsol(+$U$) gap at this $k$-point, and $E_{\mathrm{g}}^{\mathrm{hybrid,band}}$ as the minimum gap from nsc-dd hybrid calculations.

As will be shown in the Methods section, our calculations without excitonic effects or phonon-assisted electronic transitions yield the optical absorption coefficients by considering only parity-allowed direct electronic transitions from occupied bands to unoccupied bands. Therefore, with $E_{\mathrm{g}}^{\mathrm{PBEsol}(+U),\mathrm{diele}}$ denoting the minimum gap in the electronic density of states relevant to dielectric function calculations using PBEsol(+$U$), interband transitions do not occur theoretically at energies lower than $E_{\mathrm{g}}^{\mathrm{PBEsol}(+U),\mathrm{diele}} + \left(E_{\mathrm{g}}^{\mathrm{hybrid,band}} - E_{\mathrm{g}}^{\mathrm{PBEsol}(+U),\mathrm{band}}\right)$, resulting in zero values for the imaginary part of the dielectric functions and the optical absorption coefficients. Hereafter, we refer to this threshold value as the band gap. We note that $E_{\mathrm{g}}^{\mathrm{PBEsol}(+U),\mathrm{diele}}$ and $E_{\mathrm{g}}^{\mathrm{PBEsol}(+U),\mathrm{band}}$ do not strictly match because they are calculated using different $k$-points, meaning that the band gap value in this definition also does not strictly match $E_{\mathrm{g}}^{\mathrm{hybrid,band}}$.

In addition, we define the spectral onset position $\omega_{\mathrm{o}}$ as the photon energy where the spherical average of the imaginary part of the dielectric function first exceeds $\frac{5\times10^{-5}}{3}$, which is derived from the numerical precision limit of the default VASP output format. **Figure 2** shows the relationship between the band gap defined above and this spectral onset energy. The onset values of some materials are clearly larger than their band gaps. These materials have electronic structures with indirect band transitions or parity-forbidden direct transitions, which are not included in our calculations, while phonon-assisted absorption occurs in reality. Methods for calculating phonon-assisted absorption by including electron-phonon coupling have been developed.[108,110] Because of its high computational cost,[109] however, our high-throughput first-principles calculation dataset does not consider such effects.

Another contribution not considered in our calculation dataset is excitonic effects. **Figure 3** shows the comparison of theoretical and experimental[141] absorption spectra of several binary oxides and chalcogenides with direct-type band structures. As can be seen, although the corrected spectra improve the onset energies over the PBEsol results, there is a tendency to underestimate the optical absorption coefficient values in a wide energy range around the onsets, where experimental spectra exhibit sharp peaks or edges near the onsets. Among the compounds we tested, MgO shows particularly severe underestimation in the 8–12 eV range, while other materials such as $SnO_2$, ZnO, and CdS exhibit better accuracy compared to MgO. We also confirmed that the imaginary part of the dielectric function for MgO is also significantly underestimated in the 8–12 eV region. This observation is consistent with the work by Begum et al.,[142] where the dielectric function of MgO was systematically investigated using various theoretical approaches and the calculation within the independent particle approximation using the HSE06 hybrid functional[143,144] also underestimated the imaginary part of the dielectric function over a broad energy range near the onset. Their study demonstrated that this error can be substantially corrected when excitonic effects are properly included through the Bethe-Salpeter equation[145] combined with the $GW$ approximation, showing excellent agreement with experimental spectra. However, due to the high-throughput nature of our calculations, we could not employ such computationally demanding methods, as well as electron-phonon coupling effects.

## 2.2. Machine-Learning Model Construction and Prediction Accuracy

To construct the ALIGNN model of optical absorption spectra, we utilized the averaged spectra over the diagonal elements of optical absorption tensors for the training and evaluation data. The spectra were preprocessed by setting a minimum threshold of $10^2$ $cm^{-1}$ and subsequently applying a common logarithmic transformation, with Gaussian smearing applied using a standard deviation of 0.2 eV, then converted into datasets spanning 0 to 15 eV at 0.01 eV intervals (1501 data points) using linear interpolation. The upper part of **Figure 4** illustrates a schematic overview of the ALIGNN architecture. The key distinguishing feature lies in the initial construction of two distinct graph representations from the local atomic environments within the crystal structure: a crystal graph, where atoms serve as nodes and pairwise bonds function as edges, and a line graph, where pairwise bonds are represented as nodes and triplet bonds are represented as edges. These dual graph representations are subsequently processed through convolutional layers to generate output values. This approach enables the explicit representation of three-body interactions, thereby differentiating it from standard crystal graph convolutional neural networks (CGCNNs).[146] Details of the ALIGNN framework and hyperparameters we used will be provided in the Methods section.

The prediction results of the test data are shown in **Figure 5**. The histogram on the left represents the distribution of the mean absolute error (MAE). We note that 75% of the materials demonstrate good accuracy with an MAE of less than 0.14. The right panel displays predictions for four materials with the largest errors within each quartile range. For materials in the first three quartiles, ALIGNN predictions (colored lines) show good agreement with reference first-principles calculations (black lines). However, several compounds in the 4th quartile exhibit significant deviations, particularly in the onset position of the optical absorption spectra.

The poor prediction performance for these outliers can be attributed to their distinctive electronic structures and underrepresentation of similar compounds in the training dataset. As can be seen in **Figure 5**, the largest MAE in the test set is observed for $Sr_3BPO_3$ (MAE = 0.45), followed by $Ce_2SeN_2$ (MAE = 0.37) and BeO (MAE = 0.35). These materials exhibit unique density of states characteristics. While typical oxides and chalcogenides have valence bands dominated by *p*-orbital components of O or chalcogen, $Sr_3BPO_3$ shows a valence band primarily composed of P *p*-orbitals. Similarly, $Ce_2SeN_2$, which contains complex anions, exhibits significant hybridization between Ce *f*-orbitals and N *p*-orbitals near the valence band edge. For BeO, the difficulty of the prediction may be related to its extremely high onset energy (11.08 eV), which represents the highest value in the test set, while only 9 materials out of 2144 compounds in the training dataset have higher onset energies. We anticipate that strategic dataset augmentation targeting these distinctive electronic structures will reduce prediction errors for these rare material types.

### 2.3. Feature Extraction and Clustering

We performed feature extraction on the first ALIGNN layer of the optimized model and averaged the obtained feature vectors across all atomic sites for each material, followed by hierarchical clustering analysis, as schematically shown in the lower part of **Figure 4**. The objective of this approach is to classify materials into groups that exhibit similarities in both input features (such as elemental composition and atomic coordination characteristics, including the number of neighboring atoms, interatomic distances, and bond angles) and output properties (optical absorption spectra). If such material groups are successfully defined, this classification would enable us to identify common factors within these material groups and, therefore, infer the origins of desired spectral characteristics (e.g., onset values and the sharpness of absorption edges).

This methodology bears similarity to work by Kiyohara et al.,[147] where decision tree analysis was applied to interpret ELNES/XANES spectra of molecules, and our previous work, where we converted trained random forest models into *z*-vectors and performed clustering to classify material groups.[84] Employing graph neural network-based feature extraction offers three advantages over those previous works: (1) the elimination of the need to manually prepare explicit hand-crafted crystal structure features for the construction of the prediction model, as simple atomic features and interatomic distances and angles can be directly input into the neural network; (2) the ability to handle predictions involving a large number of output values such as spectral data, unlike random forests; and (3) the utilization of graph-convolutional neural networks widely used in materials science nowadays. Other deep learning-based interpretation approaches have employed dimensionality reduction techniques such as t-distributed stochastic neighbor embedding (t-SNE)[148] and uniform manifold approximation and projection (UMAP)[149] to visualize extracted feature spaces for optical spectra interpretability.[96,150] While these methods can create intuitive and accessible visual maps, they inherently involve information loss when embedding high-dimensional spaces in 2 or 3 dimensions. In contrast, our hierarchical clustering approach enables the strict definition of material classes at arbitrary granularity without information loss, and facilitates the extraction of common factors for each material class through analysis of feature distributions within clusters, as will be demonstrated later. As an approach using hierarchical clustering, Moriwake et al. performed hierarchical clustering directly on phonon dispersion curves along specific $k$-paths for materials with perovskite structures[151]. They obtained materials with phonon dispersions similar to $BaTiO_3$ and $KNbO_3$, which are widely used ferroelectric materials. Furthermore, one of these materials ($RbNbO_3$) has been experimentally demonstrated to possess a high dielectric constant. This approach is useful for directly exploring promising materials with similar properties to high-performance existing materials. On the other hand, as will be demonstrated later in this paper, since crystal structure features are not considered during classification, it becomes difficult to extract common features of constituent elements and atomic arrangements for each material group.

**Figure 6** shows the optical absorption spectra for each of the 96 groups obtained through hierarchical clustering. The spectral shapes within each cluster are indeed similar, confirming the effectiveness of our clustering approach. While some clusters with a large number of materials appear to exhibit greater variation, the hierarchical nature of the clustering allows for further subdivision into more refined groups if needed, as shown later. **Figure 7** displays scatter plots of the onset photon energy $\omega_o$ and the integrated absorption coefficient from $\omega_o$ to 0.2 eV above, defined as $I = \int_{\omega_o}^{\omega_o+0.2} \alpha(\omega)d\omega$, where $\omega$ and $\alpha(\omega)$ denote the photon energy and absorption coefficient, respectively. In these scatter plots, materials within the same cluster tend to show similar position and steepness of the absorption coefficient onset, consistent with the spectral similarities observed in **Figure 6**.

Let us take cluster 74 as an example, where the materials contained tend to exhibit relatively wide band gaps and high absorption coefficients near the spectral onset. **Figure 8** (a) shows that all materials in this cluster contain either V or Cr, with the other cations being predominantly alkali metals. Most of these materials incorporate these elements in the form of $VO_4^{3-}$, $CrO_4^{2-}$, or $Cr_2O_7^{2-}$ where the cations are in tetrahedral coordination. We calculated the tetrahedral coordination indices for the cation sites in each material within this cluster using the CrystalFingerprintNN[152] implemented in the MATMINER code[113] (version 0.9.2) and examined the distribution of the maximum values across all cation sites. As shown in **Figure 8** (b), the majority of materials contain sites with tetrahedral coordination. This observation is further confirmed by examining the representative crystal structures shown in **Figure 8** (c), where tetrahedral coordination environments are visually evident. As can be seen from their electronic densities of states, sharp peaks associated with V-*d* or Cr-*d* states are indeed observed near the conduction band minima. The high oxidation states of $V^{5+}$ and $Cr^{6+}$ provide numerous accessible unoccupied electronic states for optical transitions. Therefore, these vanadates, chromates, and dichromates exhibit high optical absorption coefficients for reasonable physical reasons from a solid-state chemistry and physics perspective. This high absorption behavior is consistent with the historical use of chromate and dichromate compounds as dyes and pigments, although such applications have been largely discontinued due to their toxicity.[153,154] On the other hand, $Rb_3V_5O_{14}$ belonging to this cluster has been investigated as environmentally friendly red fluorescent pigment[155].

Another material group including $V^{5+}$ and $Cr^{6+}$ compounds is cluster 70. As can be seen from **Figure 6** and **Figure 7**, this cluster tends to show absorption spectra with the onsets at lower photon energies compared to cluster 74. **Figure 9** shows the same information for cluster 70 as that for cluster 74 given in **Figure 8**. Unlike cluster 74, many materials in this cluster do not contain alkali metals as cations besides V and Cr. We applied the aforementioned tetrahedral coordination indices analysis as shown in **Figure 9** (b) and found that a significant number of materials in this cluster are not identified as having tetrahedral coordination. For example, $V_2O_5$ shown in **Figure 9** (c) has one type of equivalent vanadium sites with V-O bond distances of 1.60, 1.79, 1.89, 1.89, 2.01, and 2.66 Å in ascending order, where the coordination number cannot be clearly determined as four. The CrystalNN analysis yielded significant indicator scores for 5-fold or 6-fold coordination rather than tetrahedral coordination. In contrast, $HfV_2O_7$ has vanadium sites that are clearly identified as tetrahedral four-coordination. **Figure 10** shows the dendrogram at the level of 96 clusters, where clusters 74 and 70 are quite distant from each other on the dendrogram. This indicates that the prediction model considers these materials to be significantly different, despite both including vanadates and chromates. **Figure 11** (a) shows the one-electron states of the conduction band minima (CBMs) for several materials in clusters 70 and 74. In $CrO_3$, the $CrO_4$ tetrahedra form chain-like structures, and the CBM state is spread over multiple Cr and O sites. In contrast, the $CrO_4$ tetrahedra are isolated by K atoms in $K_2CrO_4$, and the CBM state shows a localized feature around these tetrahedra. Similarly, while the CBM state in $KV_3O_8$ spreads throughout the crystal, in $KVO_3$, which has a higher K content, it is localized mostly on the $VO_3$ planes interrupted by K atoms. The delocalization tendency in the materials belonging to cluster 70 would lower the CBM energy levels, resulting in lower onset energies compared to cluster 74. Regarding other surrounding cations than K, the orbital contributions from those cations can play an important role in determining the CBM characteristics. For example, the CBM state of $NbVO_5$ shows a sizable hybridization between V and Nb orbital components and is distributed throughout the crystal, in contrast to the case of $MgVBiO_5$ with a localization characteristic around V. While a rigorous discussion of the origins of onset energy and the interpretation of the prediction model based on the variety of element types and Cr/V atomic arrangements is challenging, a simplified analysis of the relationship between alkali metal density, onset energy, and cluster classification for the vanadates and chromates would be helpful, as shown in **Figure 11** (b). This plot reveals that the materials in cluster 74 tend to have higher alkali metal densities compared to those in cluster 70 and that the onset energy correlates strongly with the alkali metal density, implying that alkali metals other than K also lead to the localization of the CBM states. Furthermore, the stronger tetrahedral character observed in cluster 74 can also be attributed to the isolation of $CrO_4$ or $VO_4$ tetrahedra by surrounding alkali metal atoms, reducing hybridization of orbital components from other atomic sites and structural distortion.

Cluster 81 shows a scattered distribution of onset energies as shown in **Figure 7**, but contains materials with relatively large absorption coefficients near the onset and onset energies similar to cluster 74. The dendrogram for this cluster is shown in **Figure 12**. This cluster predominantly contains phosphorus (43 out of 50 compounds), and with the exception of $LiScP_2O_7$, all materials contain at least one of the following transition metal cations with an oxidation state of +4 or higher: $Ti^{4+}$, $V^{5+}$, $Zr^{4+}$, $Nb^{5+}$, $Mo^{6+}$, $Hf^{4+}$, or $Ta^{5+}$ and do not contain Cr. This demonstrates that our classification method can differentiate materials into separate groups even when they exhibit similar spectral features but differ in elemental composition or coordination environments. While the onset energies showed large scatter in **Figure 7**, examination of this dendrogram reveals that further subdivision could extract subclusters containing materials with onset energies similar to cluster 74. Thus, the hierarchical clustering approach employed in this study allows classification at arbitrary granularity levels, enabling more refined analysis when needed. Other clusters that appear to have broad data point distributions in **Figure 7** can also be appropriately subdivided to enable similar detailed analysis as demonstrated for cluster 70, although such comprehensive analysis is beyond the scope of this paper.

We now discuss the case where clustering was performed using the 1501-dimensional reference spectral data values instead of the intermediate features. **Figure 13** shows how chromates and dichromates in cluster 74 and 70, which were classified based on the aforementioned feature-extracted quantities, are classified when using actual reference spectral data. Although materials within the same cluster show similar optical absorption spectral shapes, the chromates and dichromates from cluster 74 and 70 are dispersed across several clusters and mixed with oxides, sulfides, and selenides having diverse elemental compositions, making it difficult to extract common characteristics for each cluster. Reference data-based clustering effectively identifies materials with similar spectra and would be useful, particularly for finding alternatives to existing materials. In contrast, our proposed classification method allows for understanding in detail how machine learning prediction models make predictions, namely extracting key factors for desired spectral shapes, and thereby providing useful physical and chemical insights for materials design.

Finally, we examine cluster 48, where materials tend to exhibit higher onset energies and smaller optical absorption coefficients near the spectral onset, compared to the previously discussed clusters. **Figure 14** presents information about cluster 48 in the same format as **Figure 8** and **Figure 9**. All materials in this cluster are oxides containing at least one of B, C, or Si. With the exception of $YAl_3(BO_3)_4$, no materials containing transition metals were found in this cluster. Similar to clusters 74 and 70, this cluster also contains cations in tetrahedral coordination. As for materials that were not identified as having tetrahedral coordination, they included threefold-coordinated B and C sites, except for $CO_2$ with linearly twofold-coordinated carbon. The lower oxidation states of $B^{3+}$, $C^{4+}$ and $Si^{4+}$, and the contributions of *s* and *p* orbitals, which are less localized than *d* orbitals, would contribute to lower densities of states near the conduction band edges, leading to the gradual absorption onsets observed in the spectra. It should be noted that ALIGNN's node features consist of one-hot encoding for each element type, with oxidation states and electronic configurations not explicitly provided as input and, therefore, this classification is a result of ALIGNN's learned representations. These results provide compelling evidence that the ALIGNN model performs appropriate classification in its initial layers based on elemental features (or possibly interatomic distances) that directly correlate with spectral characteristics. By conducting similar analyses for clusters with spectral shapes suited to each researcher's specific objectives, it should be possible to identify common features among materials that realize desired properties.

## 3. Conclusion

We constructed a first-principles calculation dataset of optical absorption spectra for 2,681 oxides, chalcogenides, and related compounds. Corrections for the spectral onset energies and shapes significantly improved agreement with reported experimental spectra over standard density functional calculations. We also developed a high-precision prediction model for optical absorption spectra using this dataset and ALIGNN. Through a combination of feature extraction and clustering analysis, we successfully extracted key elemental species and their coordination environments that dominantly determined optical absorption onset energies and intensities.

Ideally speaking, electron-hole interactions and electron-phonon coupling need to be incorporated to reproduce excitonic contributions and phonon-assisted electronic transitions, respectively, but high-throughput first-principles calculations of optical absorption spectra including such effects are computationally too demanding and have not been performed in this study. Nevertheless, our first-principles calculation dataset and the machine learning model are expected to be highly effective for future materials discovery and materials informatics research. The spectral analysis methodology proposed herein can be applied not only to optical absorption coefficients but also to other spectral properties and cases where physical and chemical properties are expressed as functions of some parameters, such as their temperature and pressure dependences, making it broadly useful for materials research.

## 4. Methods

### 4.1. High-Throughput First-Principles Calculations

Firstly, we retrieved metal oxides, chalcogenides, and related compounds that satisfy the following conditions from the Materials project database: (1) substances include at least one of O, S, and Se whose oxidation number is not necessarily −2; (2) substances neither include H, halogens, noble gases, Mn–Ni, Tc–Rh, Os–Ir, Po, lanthanoids except La and Ce, nor actinides; (3) substances do not exhibit spin-polarization; and (4) substances with space group $P1$ and/or containing more than 40 atoms in the primitive unit cell are excluded due to their high computational costs and/or ambiguities of the crystal structures. The total number of materials considered for the first-principles calculations was 9,808.

The calculations were performed using the projector augmented-wave method[156] as implemented in the VASP code[139,140,157,158] (version 5.4.4). The imaginary part of dielectric function can be computed from Fermi's golden rule and expressed as[159]

$$\varepsilon_{ab}^{\text{imag,PBE}}(\omega) = \frac{4\pi e^2}{\Omega} \lim_{q\to 0} \frac{1}{q^2} \sum_{c,v,\boldsymbol{k}} 2w_{\boldsymbol{k}} \delta(\epsilon_{c\boldsymbol{k}} - \epsilon_{v\boldsymbol{k}} - \omega) \langle u_{c\boldsymbol{k}+\boldsymbol{e}_a q} | u_{v\boldsymbol{k}} \rangle \langle u_{v\boldsymbol{k}} | u_{c\boldsymbol{k}+\boldsymbol{e}_b q} \rangle \ , \quad (1)$$

where $u_{ck}(u_{vk})$ and $\epsilon_{ck}(\epsilon_{vk})$ are the cell-periodic part of valence (conduction) band wavefunction and its eigenvalue, respectively, and $a, b$ $(= x, y, z)$ are the Cartesian directions, $\boldsymbol{k}$ is the Bloch wave vector, $e$ is the elementary charge, and Ω is the volume of the primitive unit cell. The PBEsol functional[129] with Hubbard $U$ corrections to localized states[130,131] were used for the structure optimization and the calculations of dielectric functions and formation energies. We assigned effective $U$ parameters of 3 eV to the $d$ states of Ti, V, Cr, Zr, Nb, Mo, Pd, Hf, Ta, W, Re, Pt, and Au, and 5 eV to the $d$ states of Cu, Zn, Ag, and the $f$ states of Ce.

As aforementioned, the PBEsol functional tends to severely underestimate band gaps, which affects the computed dielectric function through $\epsilon_{ck} - \epsilon_{vk}$ in Eqn. (1). Therefore, we applied a PHS (PBE+HSE06+Sum rule) method suggested by Nishiwaki and Fujiwara,[118] where the imaginary part of the dielectric function is shifted by the difference between the band gap calculated using the Heyd–Scuseria–Ernzerhof (HSE06) hybrid functional[143,144] and the PBE functional[160] ($\Delta E_{\mathrm{g}}$). During this process, the imaginary part of the dielectric function is transformed using the following **Equation** (2), to ensure that it preserves the sum rule:

$$\varepsilon_{ab}^{\mathrm{imag}}(\omega) = \frac{E - \Delta E_{\mathrm{g}}}{E}\, \varepsilon_{ab}^{\mathrm{imag,PBE}}(\omega - \Delta E_{\mathrm{g}})\,. \tag{2}$$

In this study, we employed the nsc-dd hybrid functional approach[132] instead of self-consistent HSE06 calculations to reduce computational cost and improve the band gaps of wide-gap systems,[16,132–134] thereby enabling accurate high-throughput calculations. In addition, the PBEsol functional was used instead of the PBE functional, as the former tends to yield better lattice constants for solids.[129,161,162] The electronic contributions to the dielectric constants to determine the Fock-exchange mixing parameter in the full-range hybrid functional were obtained using the random phase approximation[163] to partly compensate for the errors associated with the band gap underestimation.[109,164]

The real part of the dielectric function can be obtained by the Kramers-Kronig transformation and expressed as

$$\varepsilon_{ab}^{\mathrm{real}}(\omega) = 1 + \frac{2}{\pi} P \int_0^{\infty} \frac{\varepsilon_{ab}^{\mathrm{imag}}(\omega)}{\omega'} d\omega', \tag{3}$$

where $P$ is the principal value. Light absorption spectrum $\alpha_{ab}(\omega)$ can be calculated from $\varepsilon_{ab}^{\mathrm{real}}(\omega)$ and $\varepsilon_{ab}^{\mathrm{imag}}(\omega)$ as

$$\alpha_{ab}(\omega) = \frac{\sqrt{2}\omega}{c} \sqrt{\sqrt{\varepsilon_{ab}^{\mathrm{real}}(\omega)^2 + \varepsilon_{ab}^{\mathrm{imag}}(\omega)^2} - \varepsilon_{ab}^{\mathrm{real}}(\omega)}\,. \tag{4}$$

For the machine learning, we used averaged absorption spectra, i.e., $\alpha(\omega) = (\alpha_{xx}(\omega) + \alpha_{yy}(\omega) + \alpha_{zz}(\omega))/3$.

The correction of dielectric functions by the PHS method, the Kramers-Kronig transformation, and the calculation of absorption spectra from dielectric functions were conducted using the VISE code[12]. The SPGLIB code[165] was used to symmetrize the crystal structures and identify their space groups, and the SEEKPATH code[166] was used to determine the paths in the reciprocal spaces for the band-structure calculations.

As for the machine learning dataset, we excluded (1) materials that are unstable against competing phases in our in-house database; and (2) those with the PBEsol(+*U*) bandgaps narrower than 0.3 eV because if $\varepsilon_{ab}^{\text{imag}}(\omega)$ has non-zero values in the low $\omega$ region, Equation (3) might include a large numerical error due to the variable $\omega$ in the denominator and, therefore, the absorption spectra and the dielectric constants used for the nsc-dd hybrid functional calculations are unconvincing. The resultant number of surviving materials is 2,681.

## 4.2. Machine Learning

As described above, we employed the ALIGNN to construct the prediction model. Within the ALIGNN framework, the $i$th atomic nodes is represented by the feature vector $\boldsymbol{v}_i$. We employed one-hot encoding of atomic species for $\boldsymbol{v}_i$, following the same approach as the CGCNN. The edge features $\varepsilon_{ij}$ are defined as

$$\varepsilon_{ij} = \{\text{RBF}(\|r_{ij} - R_m\|)\}_{m=1,\ldots M_{\text{pair}}}, \tag{1}$$

where RBF(·) denotes the radial basis function, $r_{ij}$ the distance between atoms $i$ and $j$, and $R_m(m = 1, \ldots M_{\text{pair}})$ the $m$-th center of the RBF. The RBF expansion utilizes Gaussian radial basis functions as implemented in SchNet[167], converting scalar distance or angle values into feature vectors for neural network processing. The triplet features are characterized by

$$t_{ijk} = \{\text{RBF}(\|\theta_{jik} - \theta_m\|)\}_{M=1,\ldots M_{\text{triplet}}}, \tag{2}$$

where $\theta_{jik}$ is the bonding angle among atom $j$, $i$, and $k$.

The convolution operations for both ALIGNN layers and graph convolutional network (GCN) layers follow the edge-gated graph convolution formulation implemented in the open-sourced ALIGNN library (version 2024.3.24)[107], with 4 layers each for both the bond graph and line graph convolutions.

Regarding hyperparameter optimization, cross-validation is preferable for hyperparameter selection. However, due to computational constraints, we used a fixed train/validation/test (8:1:1) split, selected hyperparameters based on validation performance, and obtained the corresponding test set performance as our final results. We fixed the maximum distance of pair

RBF to be 10 Å, $M_{\mathrm{pair}}$ = 80, and $M_{\mathrm{triplet}}$= 40. Using grid search with the aforementioned split configuration, we optimized the number of nearest neighbors ($N_{\mathrm{neighbor}} = \{10, 20, 30, 40\}$) and the number of hidden features ($N_{\mathrm{h}} = \{32, 64, 128, 256\}$). As can be seen in **Figure 15**, the loss functions (MAEs) of validation and test data are not sensitive to these hyperparameters. Based on the validation performance, we selected the model with $N_{\mathrm{neighbor}} = 20$ and $N_{\mathrm{h}} = 128$, which achieved the minimum validation loss. Unless otherwise noted, all results and discussion in this article refer to this optimized model configuration. The weights of the ALIGNN model were optimized by minimizing the L1 loss function with the AdamW[168] optimizer.

Hierarchical clustering was performed using the Ward linkage method with Euclidean distance, as implemented in the SCIPY library (version 1.13.1)[169].

### Acknowledgments

This work was supported by KAKENHI (Grant Numbers: JP24K08562 and JP24H00376) from the Japan Society for the Promotion of Science, Data Creation and Utilization Type Material Research and Development Project (Grant Number: JPMXP1122683430), Design & Engineering by Joint Inverse Innovation for Materials Architecture from the Ministry of Education, Culture, Sports, Science, and Technology, and KISTEC Project. Computations were carried out using the TSUBAME4.0 supercomputer at Institute of Science Tokyo.

### Competing interests

The authors declare no conflict of interest.

### Data Availability Statement

The developed code and data developed in this work are published in https://github.com/takahashi-akira-36m/opt_absorption_alignn .

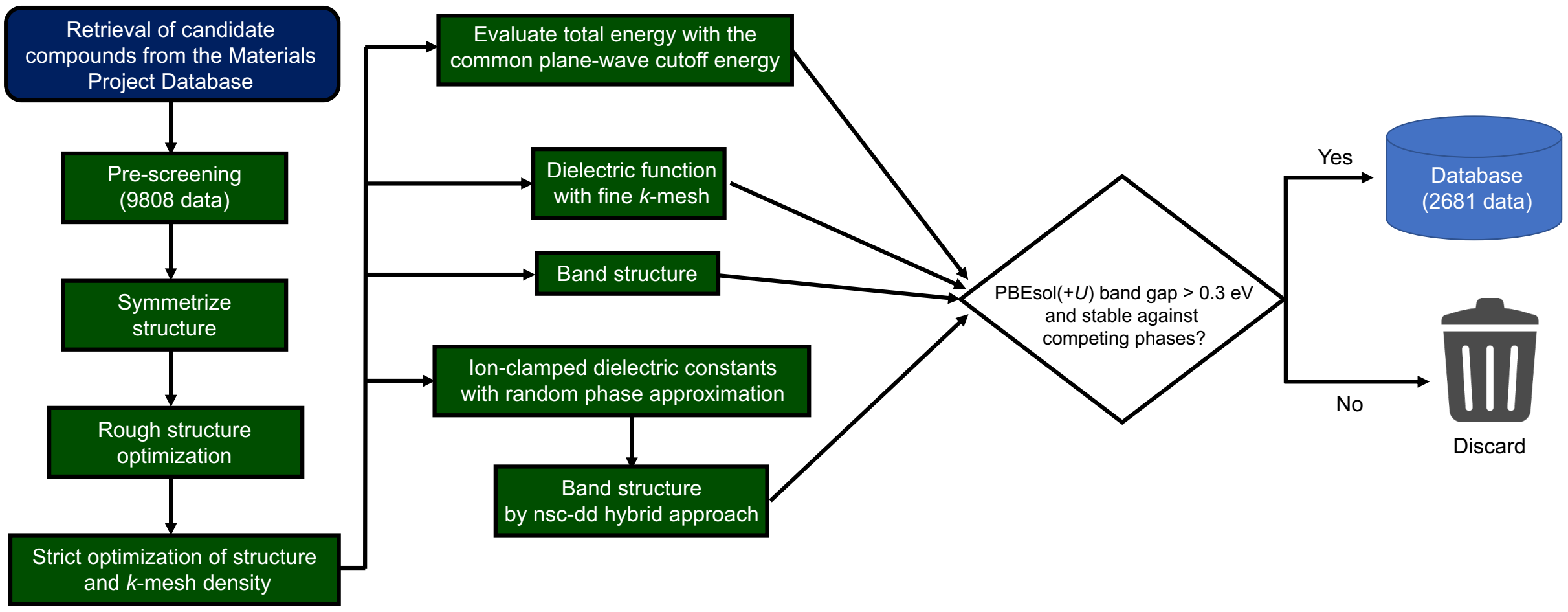

**Figure 1.** Workflow for constructing a first-principles calculation database of the dielectric functions of metal oxides, chalcogenides, and related compounds. The calculations have been performed using PBEsol(+$U$) unless otherwise noted.

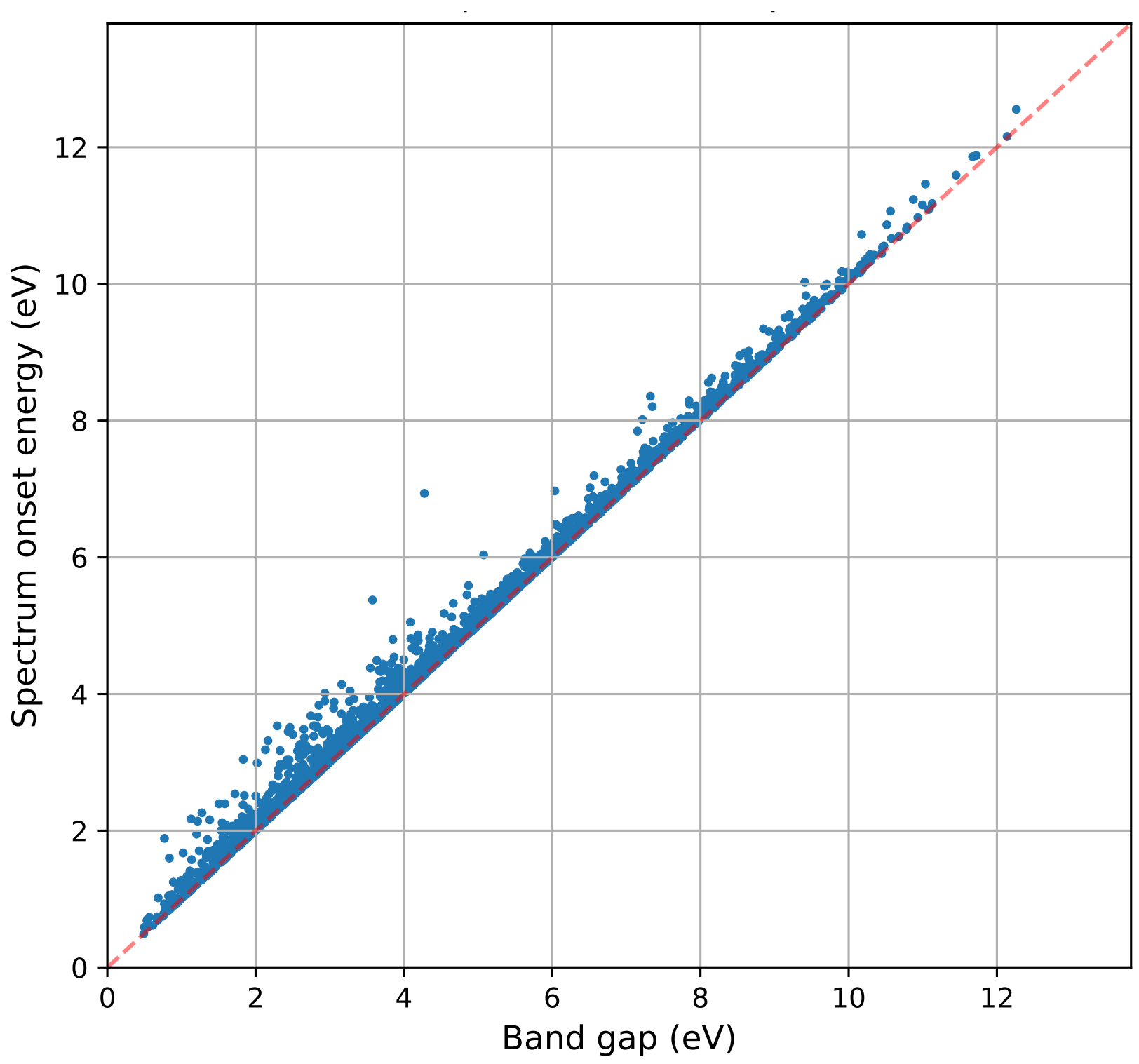

**Figure 2.** Scatter plot of the band gap and the onset photon energy of absorption spectra. The dashed red line indicates where the band gap and the onset photon energy are identical.

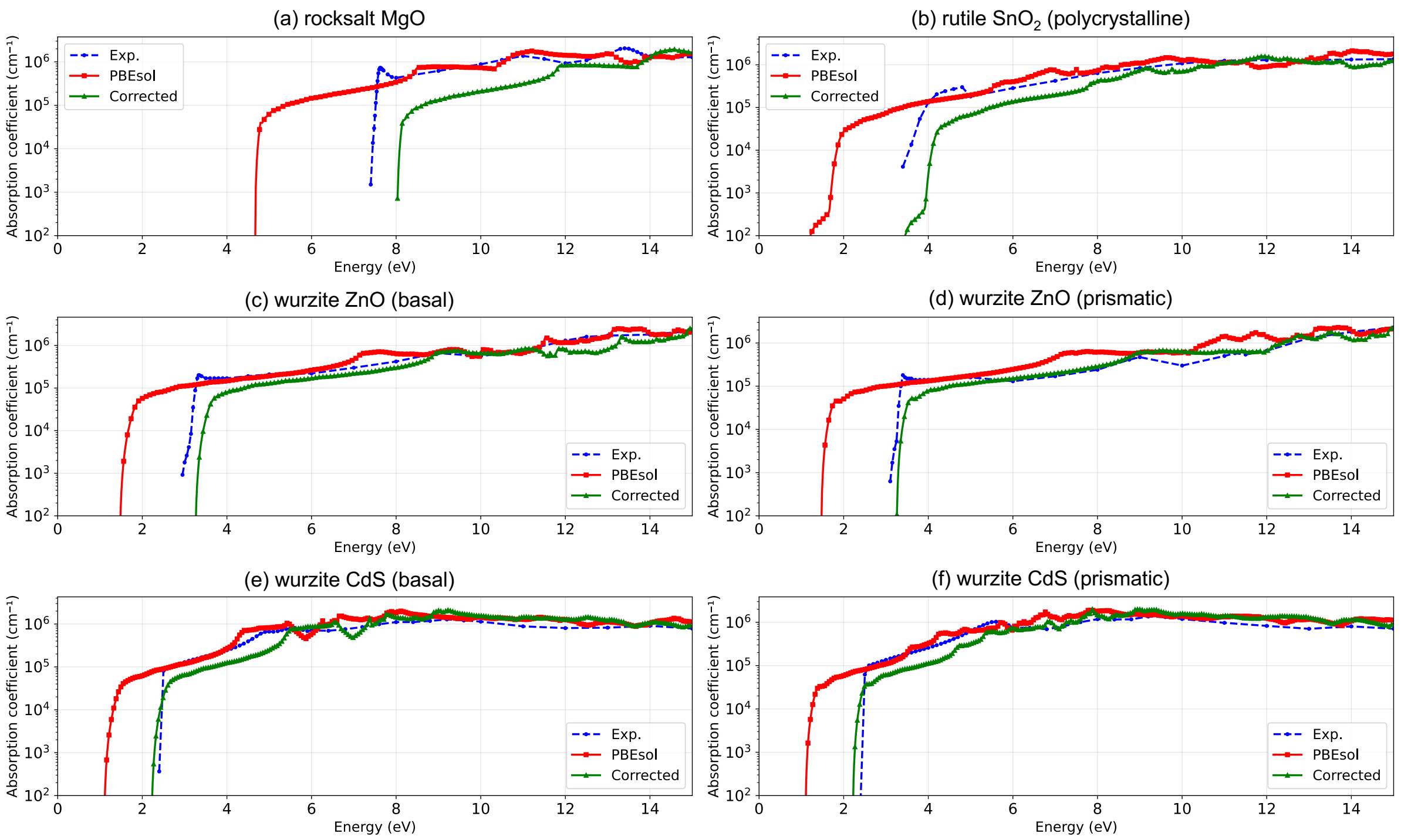

**Figure 3.** Comparison of the absorption spectra of several binary oxides and chalcogenides from experimental reports [141] (blue) and the present PBEsol calculations without (red) and with corrections (green).

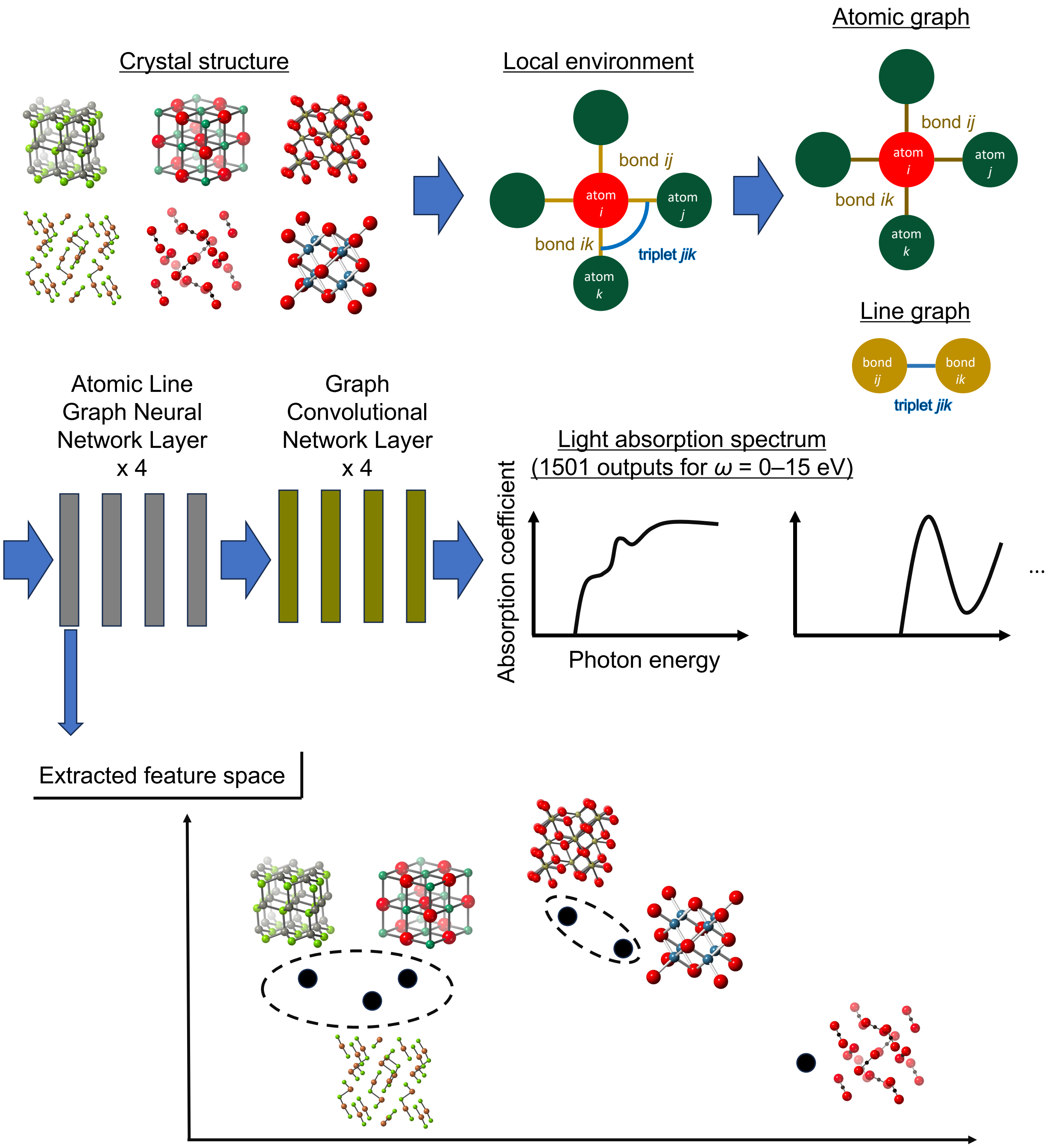

**Figure 4.** Schematic overview of our ALIGNN model for the prediction of optical absorption spectra and the proposed interpretation method.

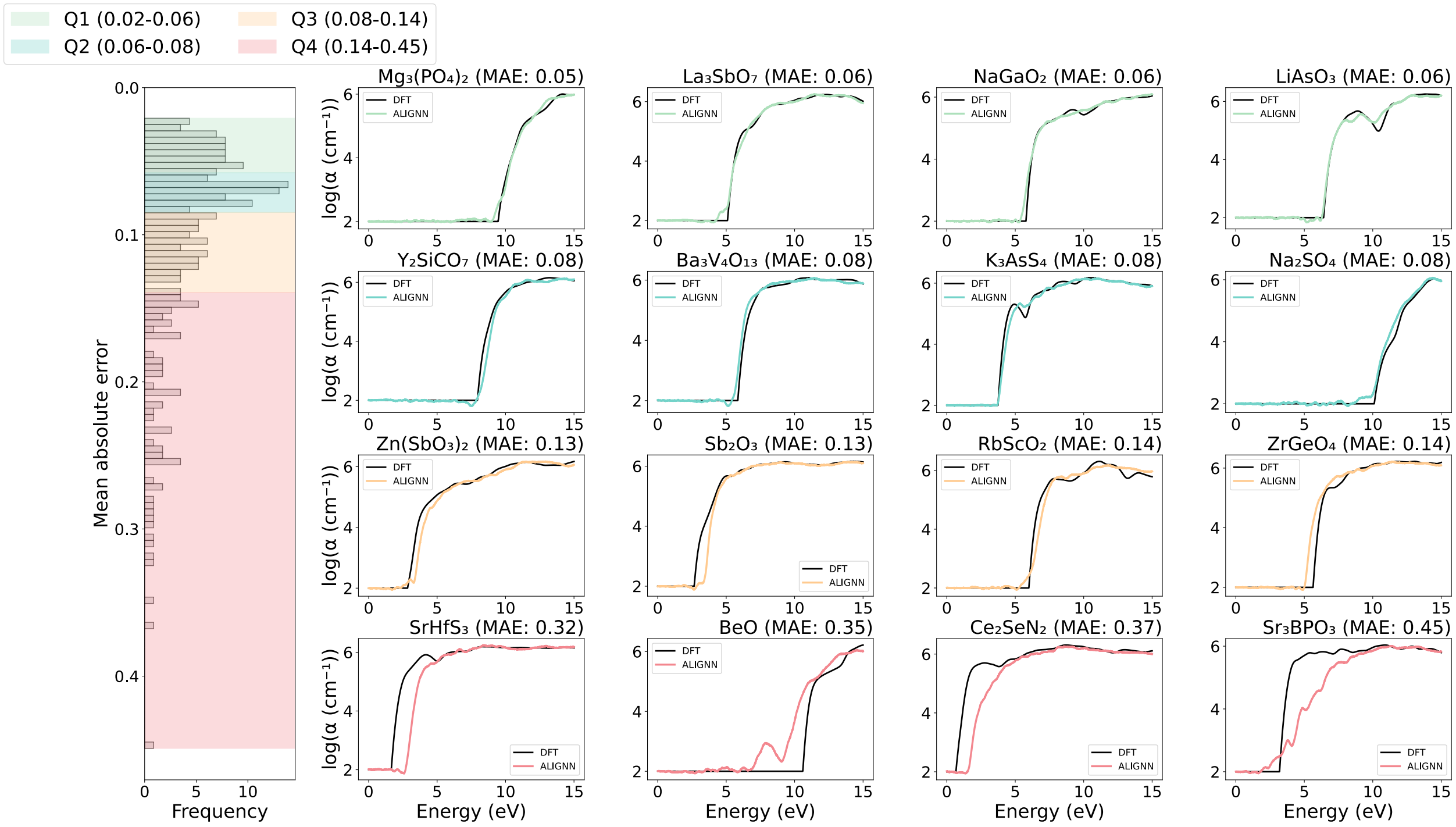

**Figure 5.** Prediction results for the test set of optical absorption spectra using the optimized ALIGNN model. (Left) Distribution of mean absolute error across test samples. (Right) Four test set examples highlighting the ones with the largest errors within each quartile.

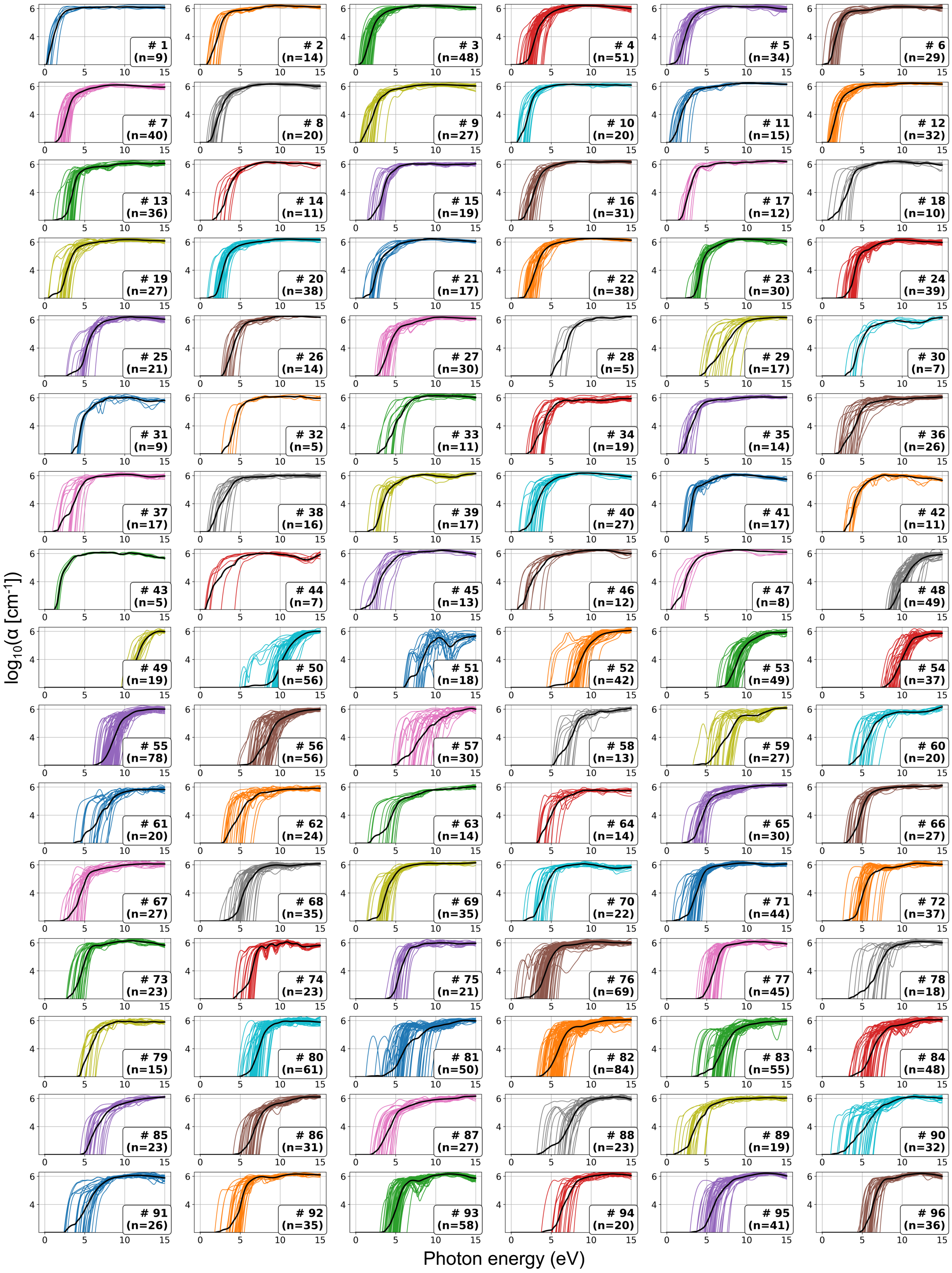

**Figure 6.** Absorption spectra classified by hierarchical clustering. Individual spectra (colored lines) and cluster means (black lines) are shown for each of the 96 clusters. The number indicates the cluster ID and *n* means the sample size.

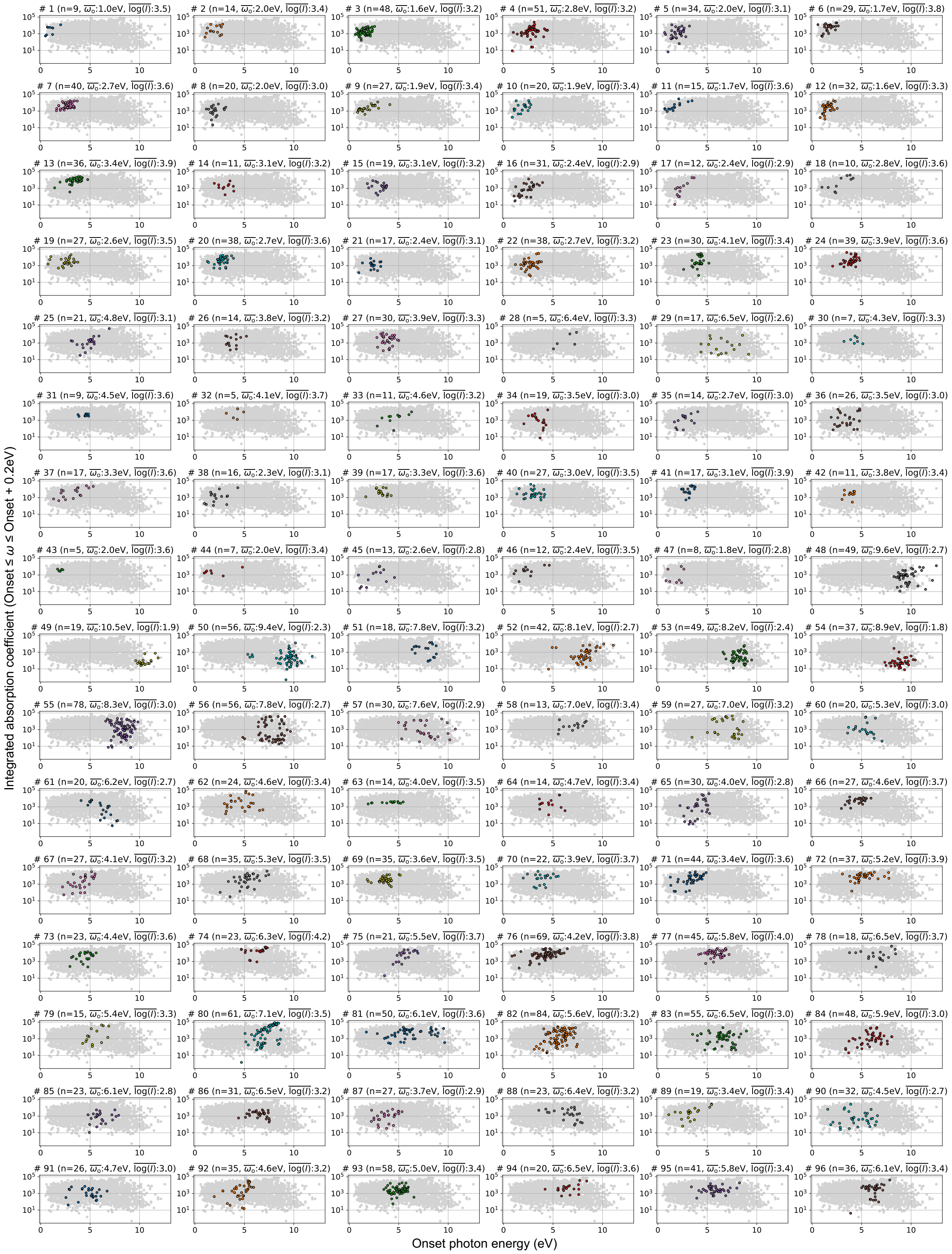

**Figure 7.** Scatter plot with the onset photon energy and the integrated absorption coefficient for materials within each cluster. $n$, $\overline{\omega_o}$, and $\overline{\log(I)}$ denote the sample size for each cluster, the mean of the onset photon energy, and the common logarithmic mean of an integrated absorption coefficient, respectively.

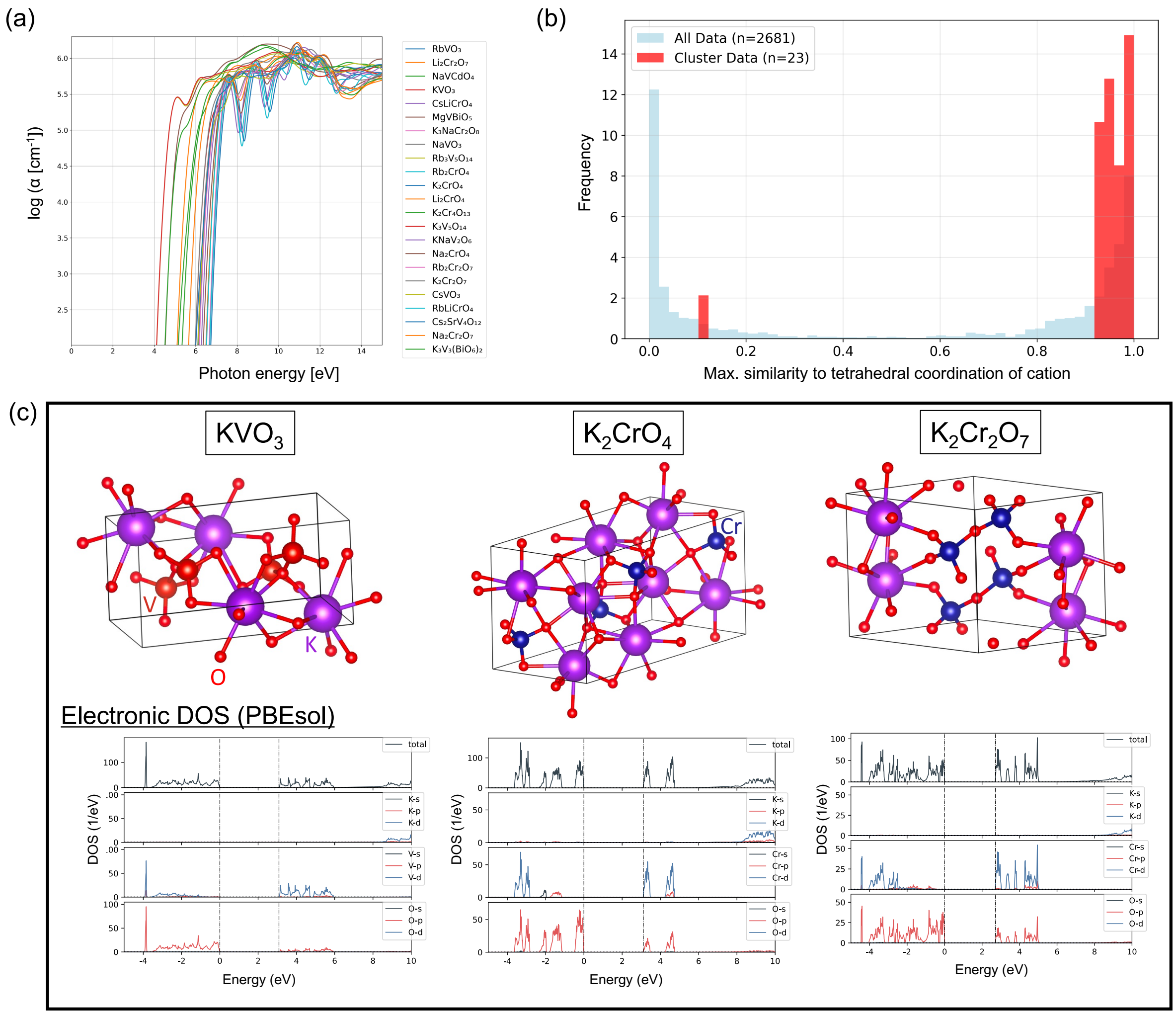

**Figure 8.** (a) Optical absorption spectra of substances belonging to cluster 74, where α denotes the absorption coefficient. (b) Distribution of tetrahedral coordination similarity for materials in cluster 74 (red) and the overall dataset (blue). (c) Crystal structures of representative materials (as drawn using the VESTA code[170]) and their electronic densities of states calculated with PBEsol (as drawn using VISE code[12]). Note that the band gaps in the displayed electronic densities of states are underestimated compared with the nsc-dd hybrid calculation results used for the spectral onset corrections.

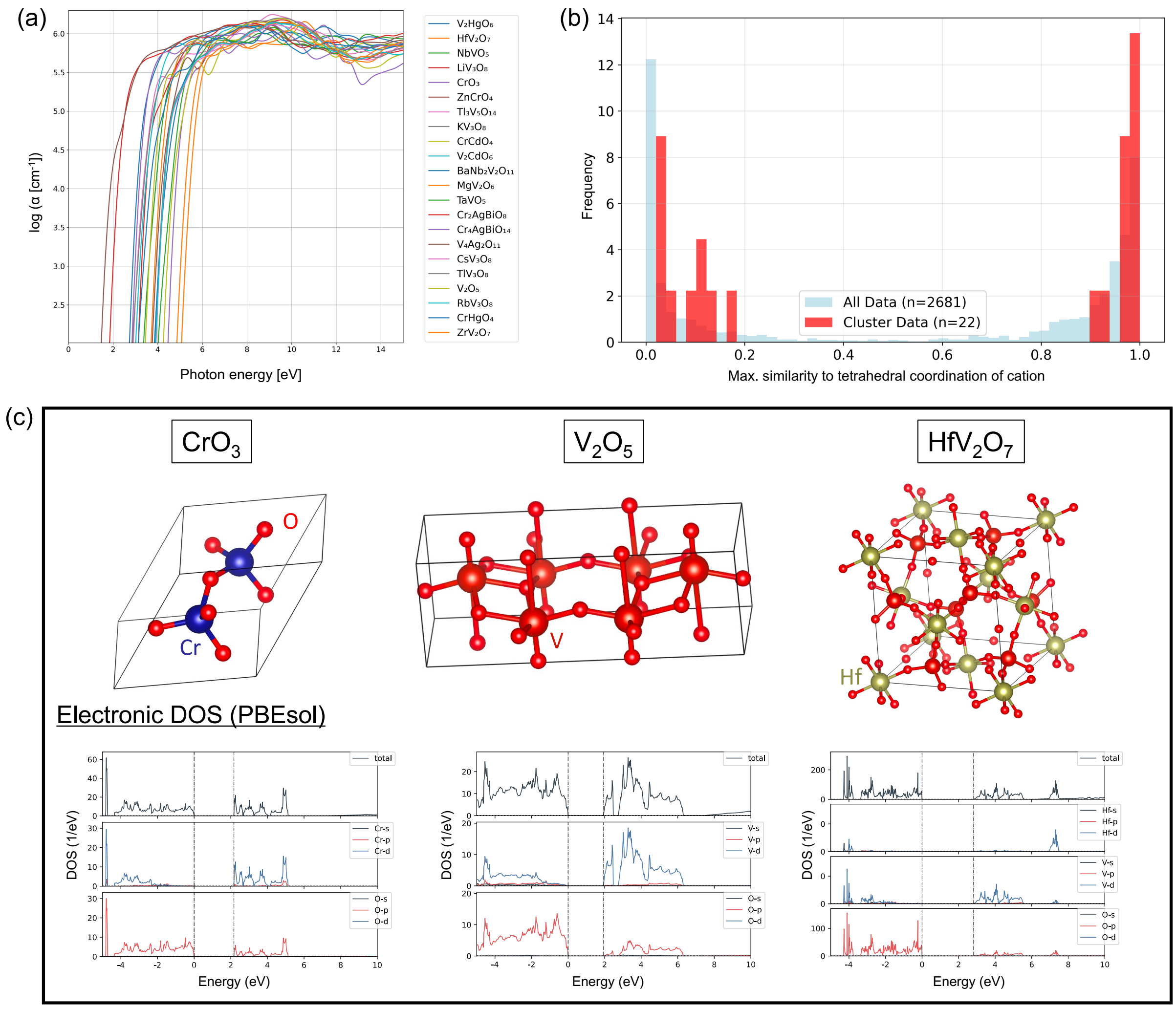

**Figure 9.** Same as **Figure 8** but for cluster 70.

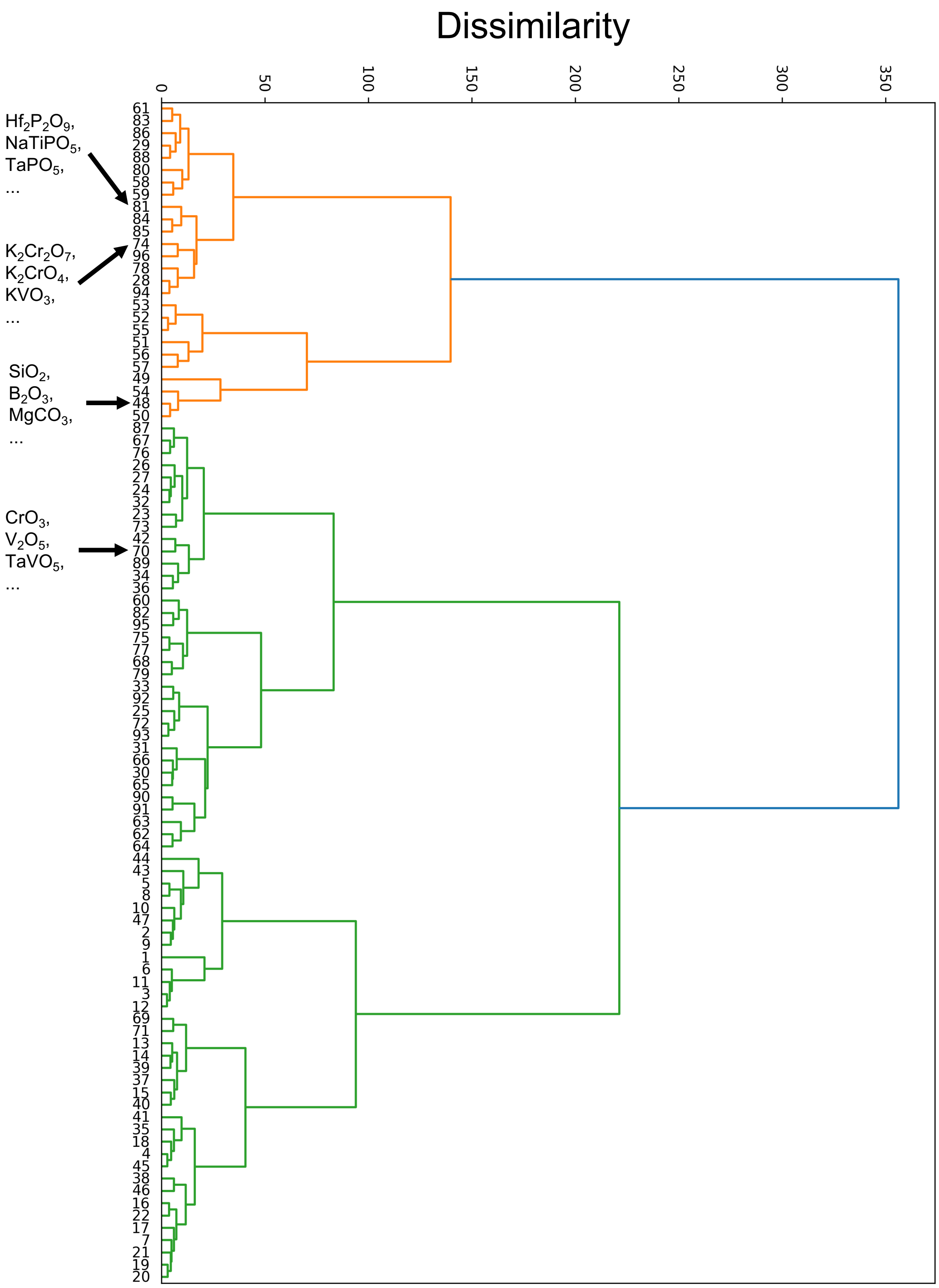

**Figure 10.** Hierarchical clustering dendrogram showing how the 96 clusters can be grouped based on their dissimilarities (Euclidean distances). This diagram groups similar clusters together by connecting them with branches. Clusters that merge at smaller dissimilarity values (closer to the left) are more similar to each other.

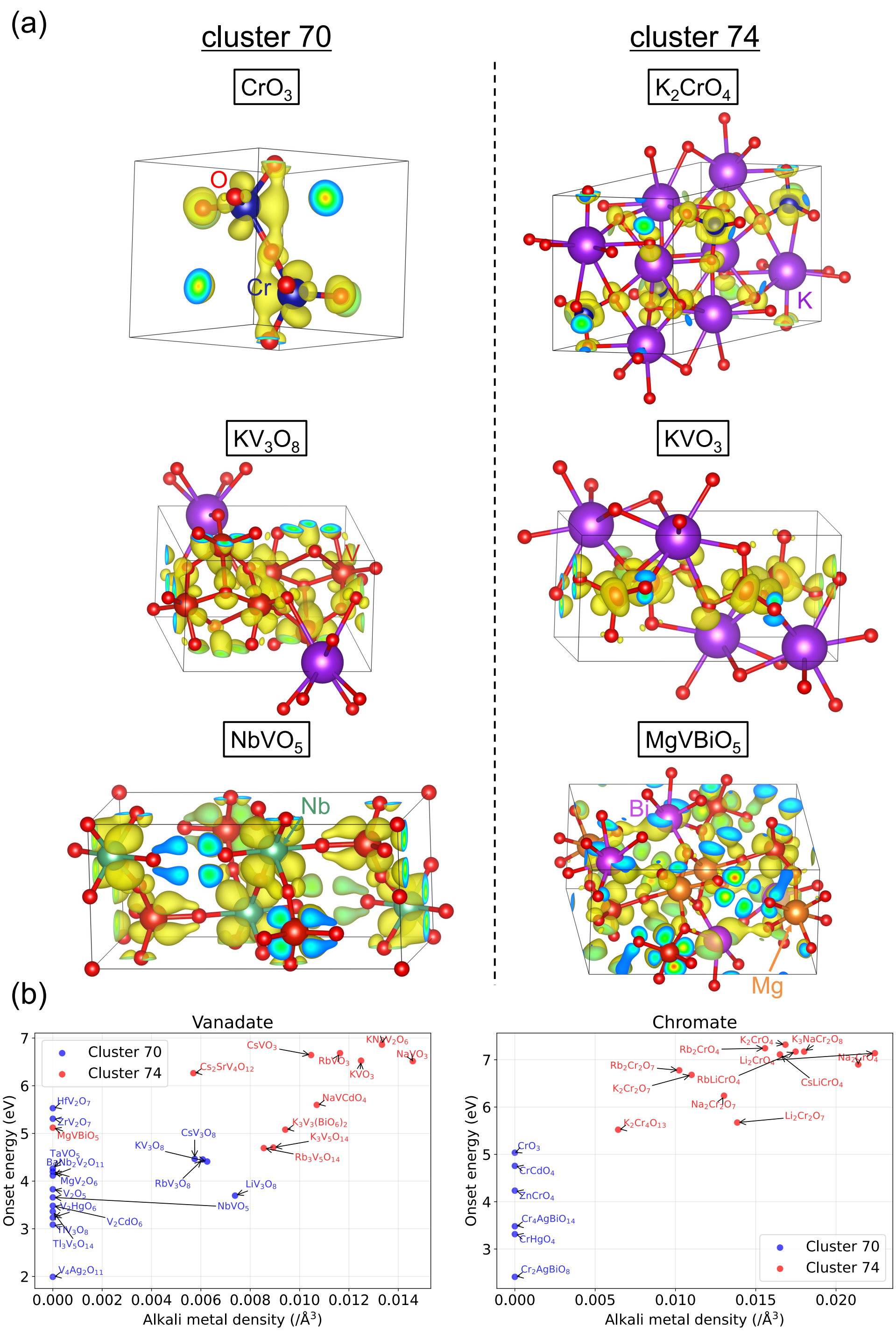

**Figure 11.** (a) Visualization of the one-electron states of the conduction band minima for several materials in clusters 70 and 74. The yellow isosurfaces are plotted at 10% of the maximum value of the squared wavefunctions of the respective materials. (b) Scatter plots showing the relationship between alkali metal density (number of atoms per unit volume) and onset energy. Wavefunctions are visualized with the VASPKIT[171] and VESTA[170] codes.

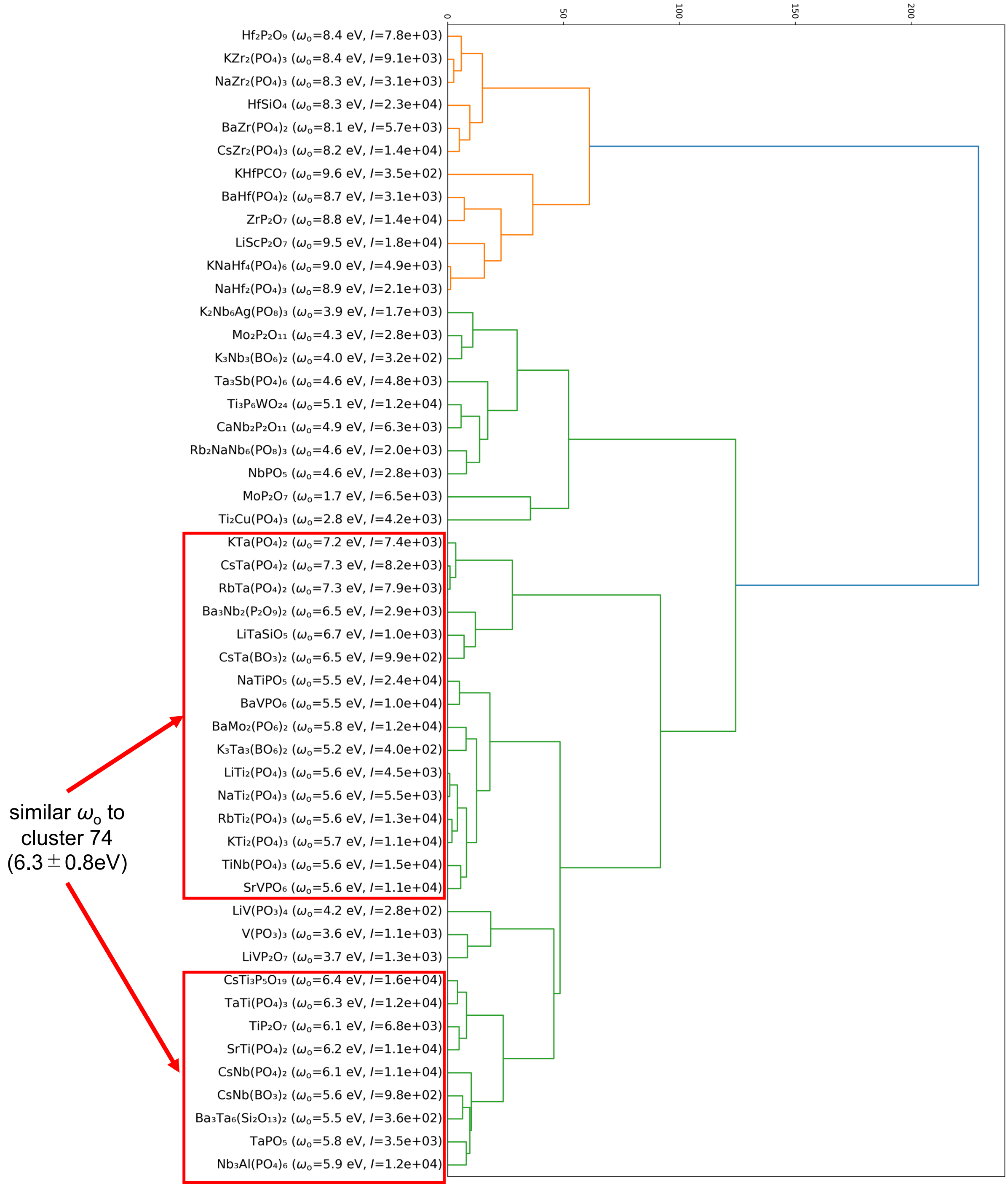

**Figure 12**. The dendrogram in cluster 81. $\omega_o$ and $I$ denote the onset photon energy and the integral of the optical absorption spectrum from $\omega_o$ to $\omega_o$ +0.2 eV, respectively.

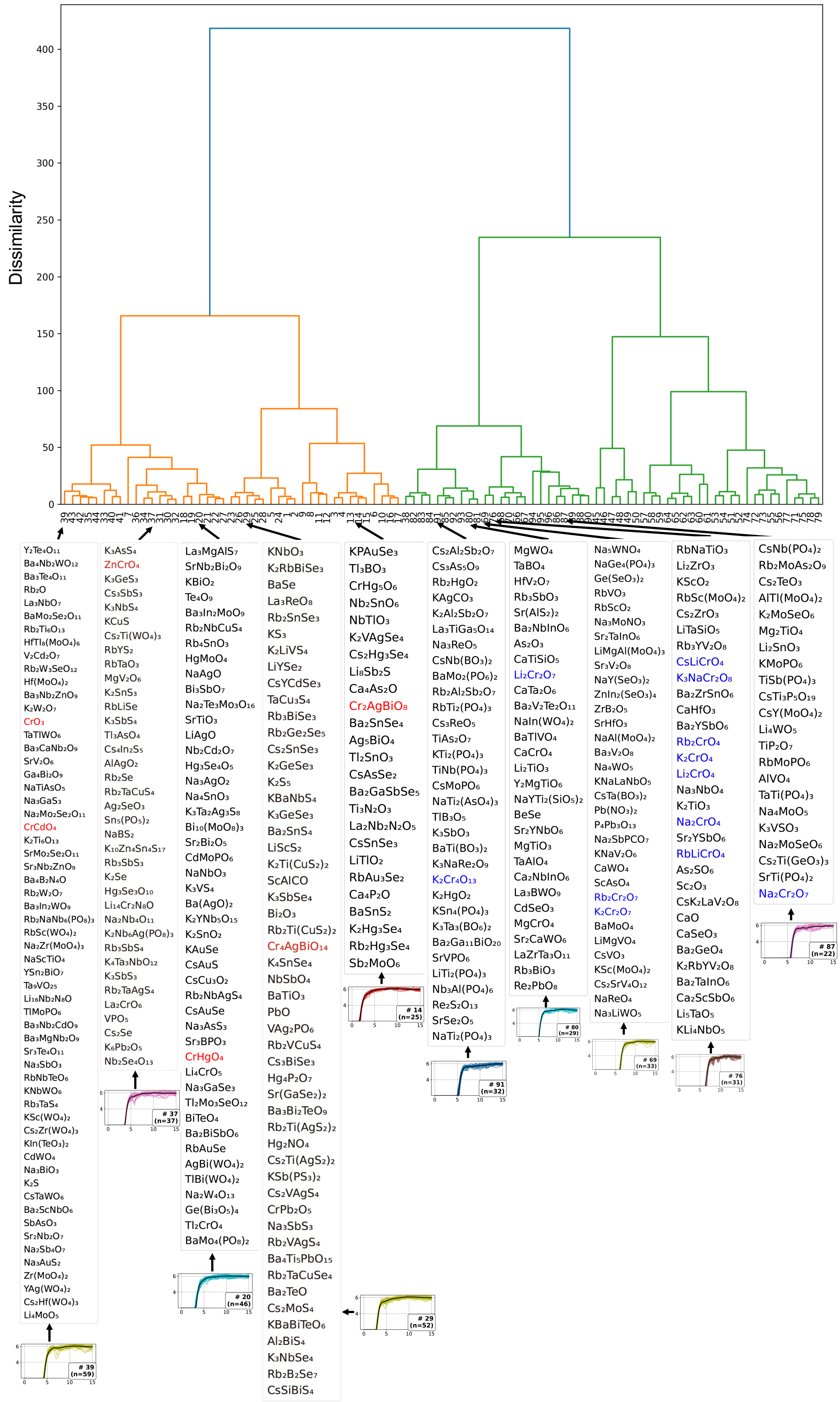

**Figure 13**. Dendrogram of clustering results using 1501 reference spectral data. The arrows indicate clusters containing chromates that belong to feature-based cluster 70 (red text) or 74 (blue text).

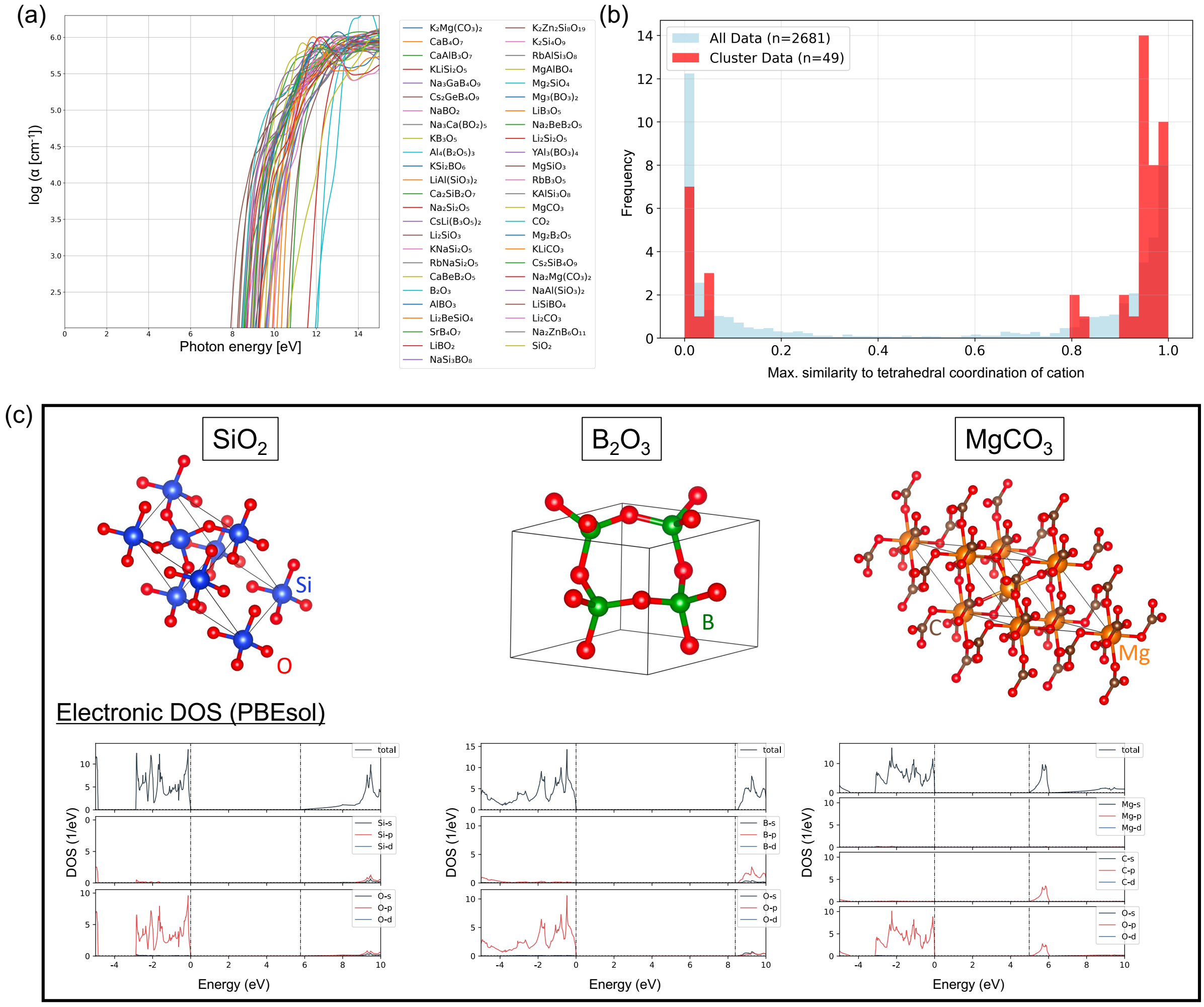

**Figure 14.** Same as **Figure 8** but for cluster 48.

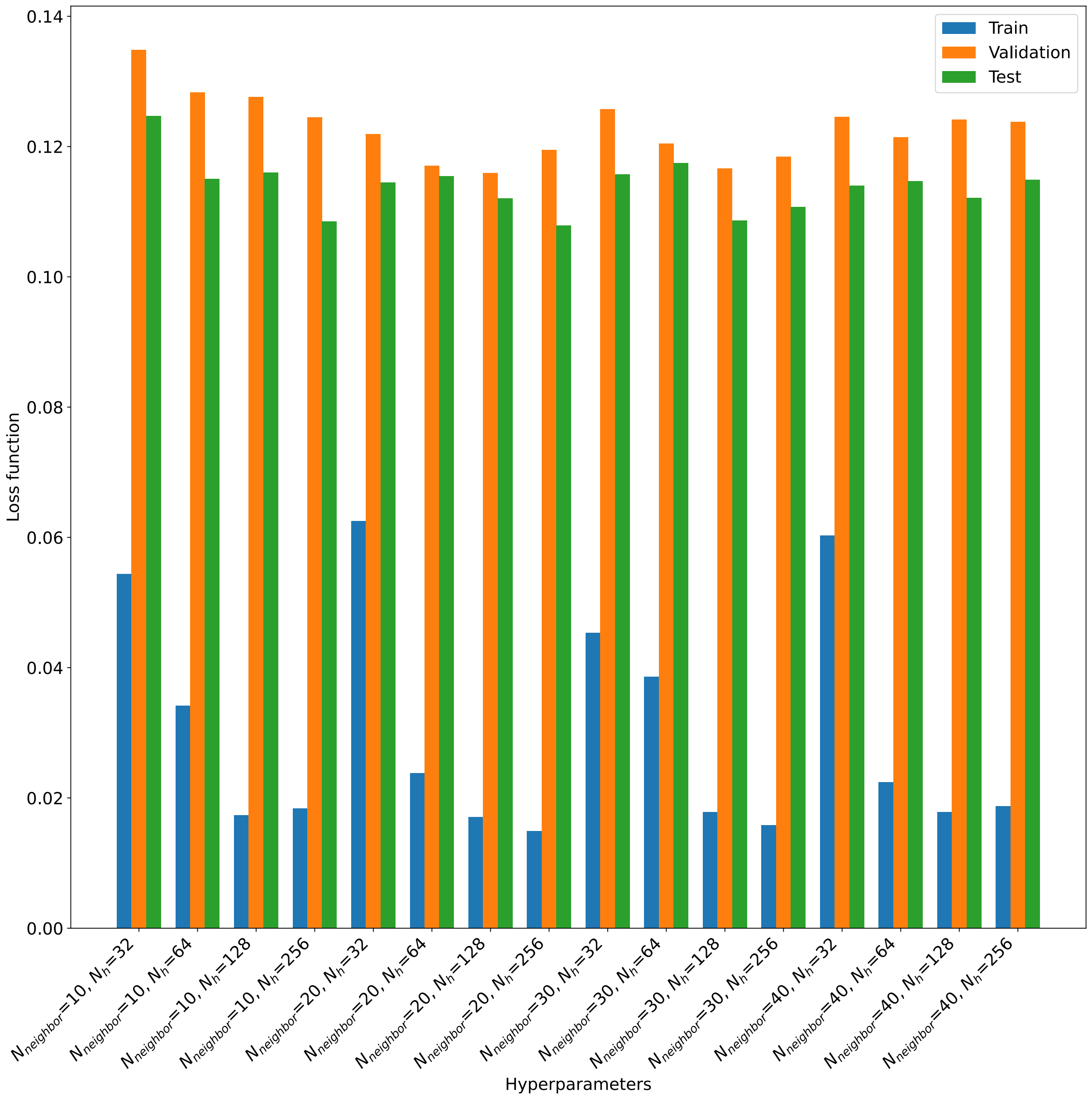

**Figure 15.** Dependency of the loss function on the number of maximum neighbors ($N_{\mathrm{neighbor}}$) and the number of hidden features ($N_{\mathrm{h}}$).